\documentclass[sn-mathphys,Numbered]{sn-jnl}% Math and Physical Sciences Reference Style
\usepackage{graphicx}%
\usepackage{multirow}%
\usepackage{amsmath,amssymb,amsfonts}%
\usepackage{amsthm}%
\usepackage{mathrsfs}%
\usepackage[title]{appendix}%
\usepackage{xcolor}%
\usepackage{textcomp}%
\usepackage{manyfoot}%
\usepackage{booktabs}%
\usepackage{algorithm}%
\usepackage{algorithmicx}%
\usepackage{algpseudocode}%
\usepackage{listings}%
\usepackage{float}
\usepackage{hyperref}
\usepackage{lineno}
\usepackage{color,soul}
\theoremstyle{thmstyleone}%
\theoremstyle{thmstyletwo}%

\theoremstyle{thmstylethree}%

\begin{document}

\title[Article Title]{Intensity-based scattering correction enables in vivo two-photon imaging beyond 1 mm}


\author[1]{\fnm{Yucheng} \sur{Li}}\email{ycsli@ucdavis.edu}
\equalcont{These authors contributed equally to this work.}

\author[1]{\fnm{Renzhi} \sur{He}}\email{cubhe@ucdavis.edu}
\equalcont{These authors contributed equally to this work.}

\author*[1]{\fnm{Yi} \sur{Xue}}\email{yxxue@ucdavis.edu}

\affil[1]{\orgdiv{Department of Biomedical Engineering}, \orgname{University of California, Davis}, \city{Davis}, \postcode{95616}, \state{CA}, \country{United States}}


\abstract{Optical imaging of the deep brain with subcellular resolution is essential for neuroscience, but noninvasive imaging beyond the cortex, through scattering white matter and into the hippocampus, has generally required three-photon microscopy at longer excitation wavelengths. Here, we introduce deep-learning-enhanced Fourier-domain intensity coupling for scattering correction (DeepFOCUS), an intensity-based two-photon approach that uses deep learning to compute intensity-modulation masks for real-time modulation of excitation light during image acquisition. Unlike deep-learning-based image restoration, this method directly improves image formation by computing intensity-modulation masks that shape the excitation light in real time, with each mask experimentally validated by the acquired fluorescence signal to avoid hallucination artifacts. Using 1035 nm excitation, we achieved in vivo two-photon imaging beyond 1 mm depth in the intact mouse brain, resolving YFP-labeled neurons and FITC-labeled blood vessels through the entire cortex and white matter down to the CA1 region of the hippocampus. DeepFOCUS extends two-photon imaging to depths previously accessible mainly with three-photon microscopy and could enable broader adoption of hippocampal imaging by upgrading existing two-photon systems.}

%%================================%%
%% Sample for structured abstract %%
%%================================%%




\maketitle

\section{Introduction}\label{sec1}

Two-photon microscopy has enabled numerous advances in neuroscience through high-resolution in vivo imaging, yet its penetration depth in the mouse brain remains largely confined to the cortex because of tissue scattering, limiting access to subcortical regions in the intact brain. State-of-the-art two-photon systems incorporating phase-modulation-based scattering correction, including adaptive optics (AO) and wavefront shaping (WS), have enabled imaging at depths of $600-800~\mu$m in the intact mouse brain using yellow-green fluorophores through cranial windows or thinned skull preparations \cite{Helmchen2005-lg, Drew2010-bp, Mittmann2011-jo, Papadopoulos2016-sy, Ji2017-vf, Takasaki2020-od, Rodriguez2021-sy, Chen2021-kq}. These approaches generally compensate for tissue-induced wavefront distortions by applying a phase-correction mask. In direct wavefront sensing and image-guided phase retrieval, the correction mask is estimated from a measured wavefront \cite{Cha2010-qe, Aviles-Espinosa2011-vt, Wang2014-wo, Wang2015-fl, Liu2019-ae, Chen2021-kq} or inferred from degraded images \cite{Paine2018-oh, Nishizaki2019-bx, Xin2019-px, Saha2020-il, Wu2020-uu, Feng2023-ln, Hu2023-jt, Kang2024-vy, Kang2026-ay}. Although effective for correcting aberration and scattering, these methods ultimately depend on ballistic photons, whose signal decays exponentially with imaging depth. Indirect wavefront sensing methods can overcome this limitation by using fluorescence feedback from multiply scattered light \cite{Albert2000-oj, Marsh2003-rh, Debarre2009-ax, Rueckel2006-lb, Tang2012-bl, Papadopoulos2016-sy, May2021-er, Rodriguez2021-sy, Qin2022-ns}. However, hardware-in-the-loop iterative optimization with phase modulators can introduce substantial latency between sensing and correction, making the correction vulnerable to animal motion during in vivo imaging. Consequently, noninvasive two-photon imaging through the entire cortex and the highly scattering white matter to reach subcortical structures such as the hippocampus remains a major challenge.

Intensity-modulation-based scattering correction has recently emerged for deep-tissue imaging with two-photon microscopy \cite{Zepeda2025-cm, He2025-au}. Rather than applying a conjugate phase to reverse wavefront distortions, this approach uses a high-speed digital micromirror device (DMD) to select beams (“phasors”) that remain approximately in phase even after multiple scattering and redistributes excitation power from out-of-phase phasors to in-phase phasors. This strategy thereby enhances constructive interference at the focus within the scattering tissue. Intensity-modulation-based scattering correction also enables faster correction than indirect phase-only modulation, not only because DMDs are typically faster than spatial light modulators, but also because it requires only a single round of sensing rather than multiple rounds of hardware-in-the-loop optimization. Previous intensity-based correction with two-photon microscopy has achieved in vivo imaging depths over 900 $\mu$m in the mouse cortex \cite{He2025-au}. However, imaging through the white matter and into subcortical regions remains difficult because the identification of in-phase phasors becomes increasingly ill-posed in highly scattering white matter. Correction masks obtained by direct summation \cite{Zepeda2025-cm} or compressive sensing \cite{He2025-au} are therefore less effective in highly scattering white matter and at greater imaging depths in subcortical regions.

Here, we introduce deep-learning-enhanced Fourier-domain intensity coupling for scattering correction, termed DeepFOCUS, an intensity-based two-photon microscopy approach that performs real-time physical scattering correction during image acquisition. DeepFOCUS shares the sensing strategy of FOCUS \cite{Zepeda2025-cm, He2025-au}: binary random intensity patterns are projected in the Fourier domain while the corresponding two-photon-excited fluorescence from the scattered focus is recorded by a photomultiplier tube (PMT) in epi-detection. The key innovation of DeepFOCUS lies in how to compute more effective correction masks from these measurements. Rather than generating the mask through direct summation \cite{Zepeda2025-cm} or compressive sensing \cite{He2025-au}, DeepFOCUS uses a lightweight self-supervised convolutional neural network (CNN), optimized directly from the acquired measurements without pretraining, to identify in-phase optical phasors and generate a binary intensity-modulation mask. The CNN-generated mask is then projected onto the DMD to physically shape the excitation light, after which the corrected focus is raster-scanned across the field-of-view (FOV) to acquire scattering-corrected images. This computational framework enables DeepFOCUS to generate effective correction masks even in highly scattering regions such as white matter.

DeepFOCUS is fundamentally distinct from deep-learning-enhanced image-guided AO and WS methods \cite{Paine2018-oh, Nishizaki2019-bx, Xin2019-px, Saha2020-il, Wu2020-uu, Hu2023-jt, Feng2023-ln, Kang2024-vy, Kang2026-ay}, in which deep-learning algorithms are applied to degraded images for post hoc restoration or to infer phase-correction masks. Such methods remain dependent on image information carried predominantly by ballistic photons. By contrast, DeepFOCUS is not image-guided but applies a CNN in the Fourier domain to directly compute intensity-correction masks in real-time from one-dimensional (1D) fluorescence measurements acquired under random binary pattern modulation. Each CNN-generated correction mask is then used to physically modulate the excitation light and is experimentally validated by the resulting fluorescence image, thereby reducing the uncertainty and risk of hallucination artifacts associated with post hoc image enhancement.

Using DeepFOCUS, we first optimized the optical and computational parameters by imaging fluorescent beads through ex vivo mouse skull. We then applied DeepFOCUS to in vivo imaging of YFP-labeled neurons and FITC-labeled blood vessels in the intact mouse brain, demonstrating imaging beyond 1 mm through the entire cortex and white matter to the CA1 region of the hippocampus. To our knowledge, this represents the deepest in vivo two-photon imaging achieved with real-time scattering correction at 1035 nm excitation, reaching a depth regime previously accessible primarily with three-photon microscopy \cite{Wang:20, Horton2013-id, Ouzounov2017-fk, Yildirim2019-wn, Streich2021-wf, Rodriguez2021-sy, Qin2022-ns} or two-photon microscopy using longer-wavelength excitation \cite{Kobat2011-ki, Kondo2017-lh, Cheng2019-zy}. DeepFOCUS thus extends the accessible depth of two-photon microscopy and provides a real-time, intensity-based scattering-correction framework for imaging across cortical and subcortical brain regions.

\section{Results}\label{sec2}
\subsection{Principle of DeepFOCUS}\label{subsec2.1}

\begin{figure}[ht]%
\centering
\includegraphics[width=1\textwidth]{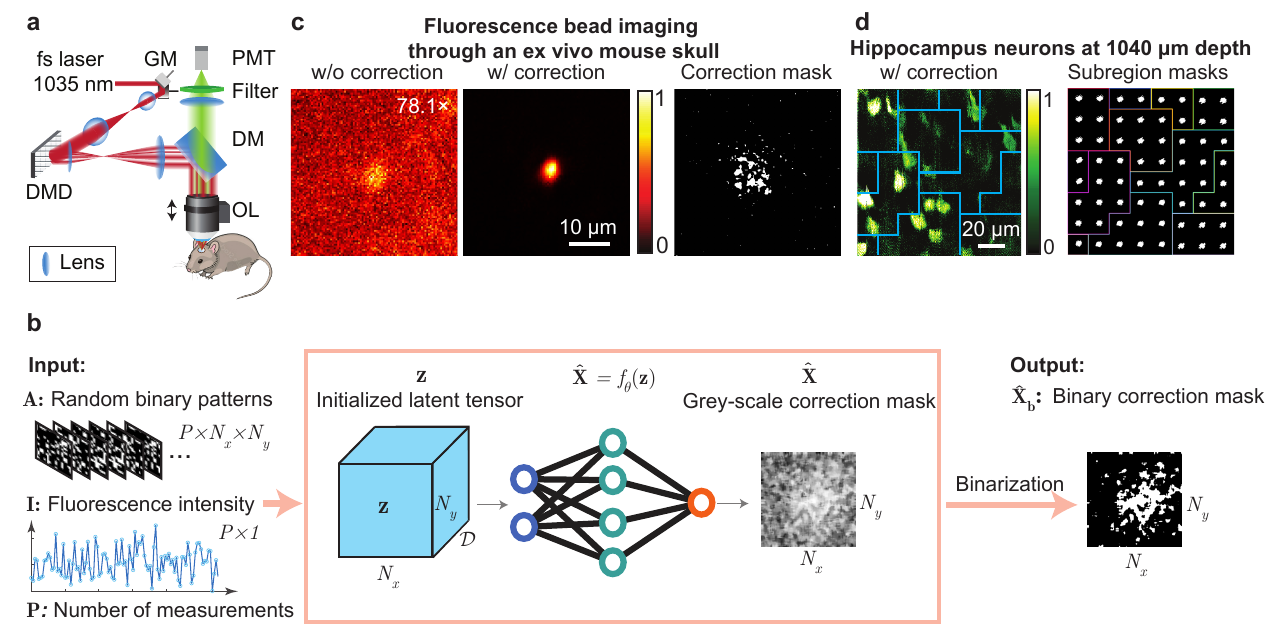}
\caption{\textbf{DeepFOCUS enables real-time, intensity-based scattering correction. a.} Optical schematic of the DeepFOCUS system. Details of the system are provided in \hyperref[Methods]{Methods}. \textbf{b.} DeepFOCUS uses a lightweight self-supervised CNN to compute the correction mask in the Fourier domain. A set of $P$ random binary patterns ($\mathbf{A}$) and the corresponding one-dimensional fluorescence measurements ($\mathbf{I}$) are used to optimize the CNN. The CNN initializes a randomly latent tensor $\mathbf{z}$ with $\mathcal{D}$ feature channels to generates a gray-scale correction mask $\hat{\mathbf{X}}$ using $k\times k$ convolution kernels. The gray-scale mask is then binarized to produce the final correction mask $\hat{\mathbf{X}}_b$. \textbf{c.} Ex vivo imaging of a fluorescence bead through a mouse skull: left, image without correction (intensity magnified 78.1-fold for display purposes); middle, image with correction; right, correction mask. \textbf{d.} Left, in vivo imaging of YFP-labeled pyramidal neurons at a depth of 1040 $\mu$m in the hippocampus of an intact mouse brain through a cranial window, with multi-region scattering correction. The subregions are indicated in the image. Right, correction masks for the individual subregions.}\label{fig1}
\end{figure}

The DeepFOCUS optical system is a point-scanning two-photon microscope with a DMD placed at a relayed pupil plane (i.e., Fourier plane) for intensity-based scattering correction (Fig.~\ref{fig1}a; see \hyperref[Methods]{Methods} for details). An excitation wavelength of 1035 nm and a repetition rate of 1 MHz were used for all experiments described below. Fluorescence images were acquired by raster scanning the excitation focus with a pair of galvanometric mirrors, while volumetric image stacks were collected by translating a high-NA objective lens (numerical aperture, 1.0) with a motorized $z$-stage. DeepFOCUS first acquired a fluorescence image without correction by scanning the unmodulated focus across the full FOV while displaying a blank mask on the DMD. Because of severe tissue scattering, the fluorescence image without correction exhibited a low signal-to-noise ratio, as illustrated by the ex vivo image of a fluorescence bead acquired through a mouse skull (Fig.~\ref{fig1}c). 

After acquiring the image without correction, we performed scattering correction with DeepFOCUS in three steps: sensing, computation, and correction (Fig.~\ref{fig1}b). DeepFOCUS addresses two key requirements for in vivo scattering correction. First, to operate beyond the ballistic-photon regime, the correction procedure must use fluorescence excited by multiply scattered excitation light as feedback. Second, to minimize the latency between sensing and correction, the entire workflow must be completed rapidly. During the sensing step, we first selected the brightest pixel in the image without correction as a target. If the FOV was larger than the optical memory-effect range, we divided the image into content-aware subregions \cite{Zepeda2025-cm, He2025-au} (Fig.~\ref{fig1}d) and selected the brightest pixel in each subregion as a target for the subsequent procedure. We parked the galvanometric mirrors at each target and displayed a sequence of $P$ random binary patterns on the DMD, each consisting of $N=N_x \times N_y$ superpixels. After flattening each two-dimensional DMD pattern into a row vector, the modulation matrix $\mathbf{A}$ can be written as
\begin{equation}
\mathbf{A} =
\begin{bmatrix}
a_{11} & a_{12} & \cdots & a_{1N} \\
a_{21} & a_{22} & \cdots & a_{2N} \\
\vdots & \vdots & \ddots & \vdots \\
a_{P1} & a_{P2} & \cdots & a_{PN}
\end{bmatrix}
\in \{0,1\}^{P \times N},
\end{equation}
where $a_{pj}$ denotes whether the $j$-th DMD superpixel is turned on in the $p$-th random pattern. For each projected pattern, DeepFOCUS recorded the corresponding two-photon-excited fluorescence intensity at the target using a PMT in epi-detection mode. The measured fluorescence response $\mathbf{I}$ is a 1D vector represented as
\begin{equation}
\mathbf{I} =
\begin{bmatrix}
I_1 & I_2 & \cdots & I_P
\end{bmatrix}^{T}
\in \mathbb{R}^{P \times 1}.
\end{equation}
Therefore, the forward model of the sensing process can be expressed as
\begin{equation}
\mathbf{I} = \mathbf{A x} + \boldsymbol{\epsilon},
\label{eq_forward}
\end{equation}
where $\boldsymbol{\epsilon}$ represents measurement noise, and
\begin{equation}
\mathbf{x} =
\begin{bmatrix}
x_1 & x_2 & \cdots & x_N
\end{bmatrix}^{T}
\in \mathbb{R}^{N \times 1}
\end{equation}
represents the desired gray-scale intensity-correction mask. At a speckle with nonzero intensity, the vector sum of all phasors defines a nonzero global phasor. Consequently, the phase distribution of the individual phasors is slightly biased toward the phase of this global phasor, such that there is an excess of phasors approximately aligned with it. Random binary patterns uniformly sample subsets of these phasors. Patterns containing a larger fraction of phasors aligned with the global phasor tend to produce stronger constructive interference and therefore higher fluorescence intensity. As a result, phasors that are approximately in phase with the global phasor are statistically overrepresented among the random patterns associated with high fluorescence signals and can be identified by large positive entries of $\mathbf{x}$ \cite{He2025-au}. Estimating $\mathbf{x}$ therefore allows DeepFOCUS to identify the phasors that should be selected and turned on in the final binary correction mask $\mathbf{x}_b$ (details are provided below). This sensing strategy uses two-photon-excited fluorescence generated by the interference of multiply scattered excitation light as feedback rather than relying on spatial information from an image, thereby bypassing the ballistic-photon limitation. Because the sensing step records only a 1D fluorescence vector rather than a 2D or 3D fluorescence image, it provides a rapid feedback mechanism for scattering correction in deep tissue. The typical exposure time for each random pattern was 0.5--2 ms; therefore, the total sensing time was determined primarily by the number of measurements $P$ and was approximately 5--18 s per correction mask in our experiments (Supplementary Table~\ref{TableS1}--\ref{TableS2}).

After sensing, DeepFOCUS computed the binary correction masks from the random binary patterns $\mathbf{A}$ and the measured fluorescence responses $\mathbf{I}$ in parallel for all subregions. The computational method was designed to satisfy three requirements. First, it should adapt to previously unseen regions with unknown scattering properties without pretraining. Second, it should remain robust to measurement noise despite the limited number of fluorescence measurements ($P$). Third, it should be computationally efficient enough to enable real-time scattering correction. To meet these requirements, DeepFOCUS uses a self-supervised lightweight CNN to estimate the gray-scale correction mask $\mathbf{x}$ directly from each experimentally measured pair $(\mathbf{A},\mathbf{I})$. The self-supervised CNN is not trained using an external training dataset. Instead, it adopts an untrained-network parameterization inspired by the deep image prior \cite{Ulyanov_2018_CVPR}, in which the estimated gray-scale correction mask is parameterized as the output of the lightweight CNN,
\begin{equation}
\hat{\mathbf{X}} = f_{\theta}(\mathbf{z}),
\end{equation}
where $f_{\theta}$ denotes the lightweight CNN, $\theta$ denotes its convolutional weights and biases, $\hat{\mathbf{X}}$ is the 2D representation of $\hat{\mathbf{x}}$, and $\mathbf{z}$ denotes a latent feature tensor with $\mathcal{D}$ channels (i.e., the feature dimension; Fig.~\ref{fig1}b). Specifically, $\mathbf{z}$ consists of $\mathcal{D}$ 2D latent feature maps $\mathbf{Z}_c$, for $c=1,\ldots,\mathcal{D}$. Both $\theta$ and $\mathbf{z}$ are randomly initialized and optimized jointly for each correction target; no parameters are learned from external training data. The network applies $L$ convolutional layers to $\mathbf{z}$. For the single-layer configuration ($L=1$) used in our experiments, the predicted 2D correction mask is
\begin{equation}
\hat{\mathbf{X}}=
\sum_{c=1}^{\mathcal{D}}
\mathbf{K}_c \ast \mathbf{Z}_c + b,
\end{equation}
where $\mathbf{K}_c$ is the $k\times k$ convolution kernel, $\ast$ denotes 2D convolution, and $b$ is a scalar bias. Each latent feature map $\mathbf{Z}_c$ has dimensions $(N_x+k-1)\times(N_y+k-1)$. Because the convolution is performed without zero padding, convolution with a $k\times k$ kernel produces an $N_x\times N_y$ output, such that the resulting correction mask $\hat{\mathbf{X}}$ exactly matches the DMD superpixel grid. The single convolutional layer contains $k^2\mathcal{D}$ kernel coefficients and one bias parameter, in addition to the trainable latent variables in $\mathbf{z}$. The predicted fluorescence measurements under the random patterns modulation is then
\begin{equation}
\hat{\mathbf{I}} = \mathbf{A}\hat{\mathbf{x}}.
\end{equation}
The network parameters and latent variables are jointly optimized by minimizing the discrepancy between the measured and predicted fluorescence measurements,
\begin{equation}
\mathcal{L}(\theta,\mathbf{z})
=
\left\|
\hat{\mathbf{I}}-\mathbf{I}
\right\|_{2}^{2}
+
\lambda\,\mathcal{R}\!\left(\hat{\mathbf{X}}\right),
\end{equation}
where $\mathcal{R}$ denotes the total variation (TV) regularizer applied to the 2D predicted mask $\hat{\mathbf{X}}$, and $\lambda$ controls the strength of the regularization. This self-supervised loss allows both the network parameters and latent variables to be optimized directly from the fluorescence measurements acquired at each target, without requiring a pretrained model or a ground-truth correction mask. Beyond the explicit TV regularization, the untrained-network parameterization itself provides implicit regularization \cite{Ulyanov_2018_CVPR, Heckel2019-dd, Heckel2020-cs}, biasing gradient-based optimization toward spatially structured solutions of the underdetermined system (Eq.~\ref{eq_forward}) and improving robustness to measurement noise compared with direct inversion. At the same time, the lightweight network architecture keeps the optimization computationally efficient, enabling real-time mask computation during image acquisition. The typical computation time per correction mask was 5--7 s on a single GPU (RTX 4000 Ada, NVIDIA), including the overhead associated with software-environment initialization and data loading (details of the computational settings are provided in \hyperref[Methods]{Methods}).

Next, the estimated gray-scale correction mask $\hat{\mathbf{x}}$ is binarized to obtain the final binary correction mask $\hat{\mathbf{x}}_b$:
\begin{equation} 
\hat{x}_{b,j} = \begin{cases} 1, & \hat{x}_j \geq T, \\ 0, & \hat{x}_j < T, \end{cases} \qquad j=1,\ldots,N, \end{equation}
where $T$ is the binarization threshold. The threshold selects the DMD superpixels with the largest estimated constructive contributions to the target fluorescence signal. It also determines the fraction of DMD superpixels that are turned on and thereby affects the effective numerical aperture, image resolution, and contrast. In practice, $T$ was determined empirically by adjusting the DMD output-to-input power ratio, with optimal values typically ranging from 0.1 to 0.4 depending on the desired balance among correction efficiency, resolution, and contrast \cite{Zepeda2025-cm, He2025-au}. 

Finally, DeepFOCUS projected the CNN-generated correction mask $\hat{\mathbf{x}}_b$ onto the DMD to physically shape the excitation light while increasing the laser power incident on the DMD to maintain the same laser power at the sample surface. DeepFOCUS then scanned the corrected focus across the entire FOV to acquire the corrected image, resulting in a substantial increase in fluorescence intensity, as illustrated by the image of a fluorescent bead (e.g., a 78.1-fold increase, Fig.~\ref{fig1}c). For multi-region correction, DeepFOCUS scanned the corrected focus while synchronously switching among the correction masks corresponding to the individual subregions, as demonstrated by the in vivo image of YFP-labeled neurons in the hippocampus of an intact mouse brain (Fig.~\ref{fig1}d). This workflow enables real-time, multi-region scattering correction across a large FOV, overcoming the conventional scattering-limited imaging depth and enabling in vivo two-photon imaging beyond 1 mm in the mouse hippocampus.

\subsection{Quantitative evaluation and optimization of DeepFOCUS parameters}\label{subsec2.2}

We quantitatively evaluated how key physical and network parameters influence the scattering-correction performance of DeepFOCUS. We imaged red fluorescence beads embedded in PDMS through a piece of 250 $\mu$m-thick mouse skull (approximately 5.3 effective attenuation lengths (EALs) \cite{He2025-au}). Details of sample preparation and imaging parameters are provided in \hyperref[Methods]{Methods} and Supplementary Table~\ref{TableS1}. Each parameter was evaluated in an independent experiment using the optimization procedure described above to empirically determine the parameters used in subsequent experiments.

\begin{figure}[H]%
\centering
\includegraphics[width=1\textwidth]{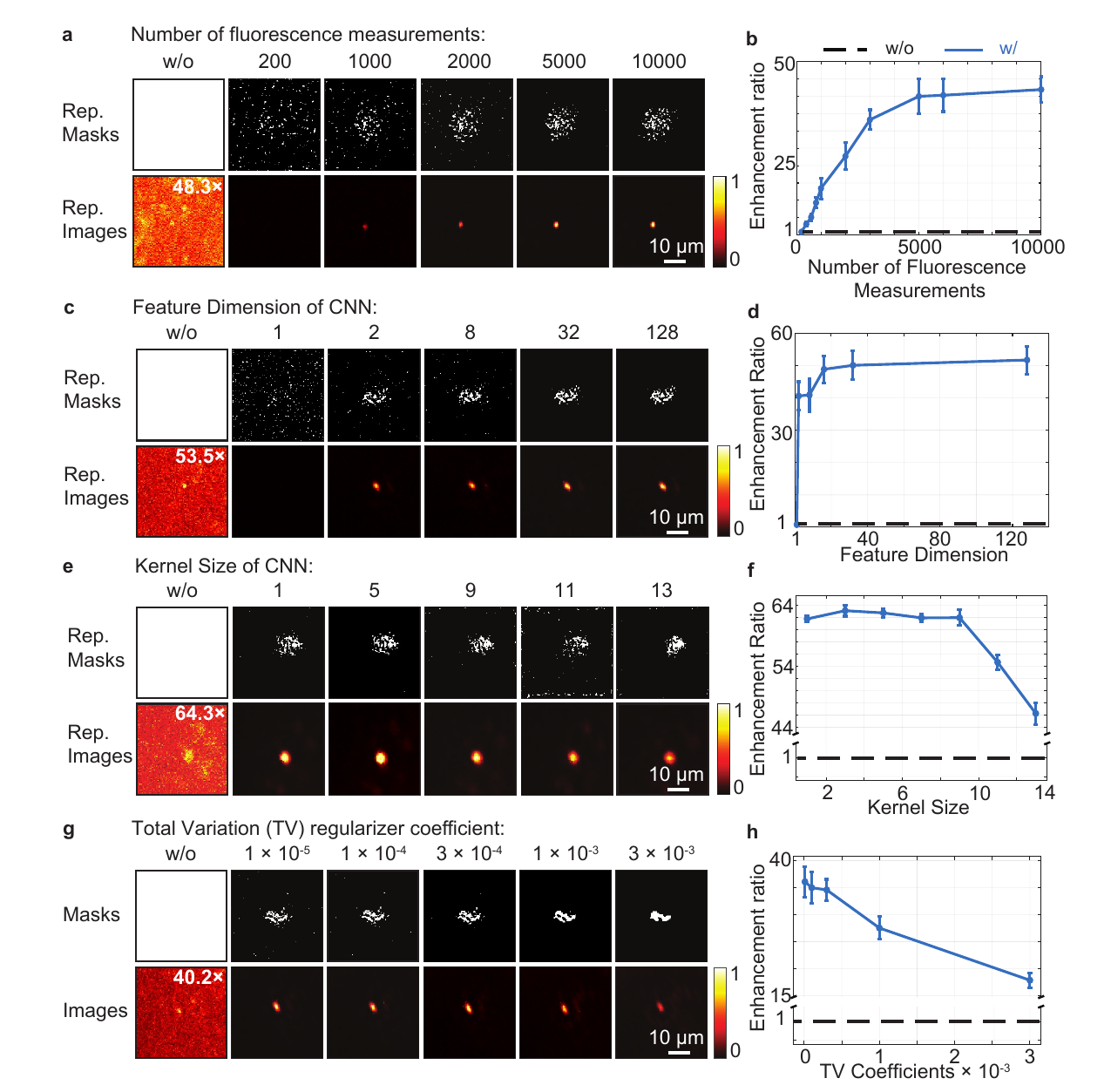}
\caption{\textbf{Quantitative evaluation of DeepFOCUS using red fluorescent beads imaged through a 250-$\mu$m-thick mouse skull ex vivo.} \textbf{a,c,e,g.} Representative fluorescence images acquired without correction and with correction while varying the number of measurements (\textbf{a}, 200--10,000), CNN feature dimension (\textbf{c}, 1--128), kernel size (\textbf{e}, 1--13), or TV regularization coefficient (\textbf{g}, $1\times10^{-5}$--$3\times10^{-3}$), together with the corresponding correction masks. For visualization, the uncorrected images in \textbf{a,c,e,g} are displayed with intensities scaled by 48.3-, 53.5-, 64.3-, and 40.2-fold, respectively. \textbf{b,d,f,h.} Fluorescence intensity enhancement ratios as a function of the corresponding parameters varied in \textbf{a,c,e,g}, respectively. Enhancement ratios are calculated from the mean intensity of the brightest 0.1\% of pixels in the in-focus plane of each fluorescence $z$-stack and normalized to the corresponding fluorescence intensity without correction. Error bars denote the standard deviation of these pixels in the same normalized units. Additional fluorescence images and correction masks are shown in Supplementary Figure~\ref{figS1}.}\label{fig2}
\end{figure}

We first investigated an essential physical parameter in the sensing process: the number of fluorescence measurements used to generate a correction mask (Fig.~\ref{fig2}a-b). This parameter directly affects both the performance and speed of scattering correction. Because DeepFOCUS estimates the correction mask from an undersampled inverse problem, increasing the number of measurements can improve estimation accuracy but also increases acquisition time and computational cost. We first imaged the fluorescent beads without correction by displaying a blank mask on the DMD. We then generated correction masks using between 200 and 10,000 fluorescence measurements and re-imaged the same beads with the corresponding masks. The fluorescence enhancement increased with the number of measurements (Fig.~\ref{fig2}b), reaching a maximum of 42-fold with 10,000 measurements. The enhancement began to plateau beyond 5,000 measurements, at which point an enhancement of more than 40-fold was achieved. Based on these results, 5,000 measurements were sufficient to reliably achieve effective scattering correction and were therefore used in subsequent experiments.

Next, we evaluated the feature dimension $\mathcal{D}$, defined as the number of channels in the latent feature tensor $\mathbf{z}$. The feature dimension controls the degree of overparameterization and representational redundancy in the mask parameterization. A small feature dimension restricts the number of trainable latent features and may make the estimated mask more sensitive to initialization and measurement noise, whereas increasing $\mathcal{D}$ provides additional degrees of freedom that can improve optimization stability and robustness \cite{Geiger2020-sc}. Beyond a certain degree of overparameterization, however, the performance gain is expected to diminish \cite{Jacot2018-ntk}, while computational cost and memory consumption continue to increase. We therefore evaluated feature dimensions ranging from 1 to 128 using the same set of fluorescence measurements to generate correction masks and subsequently acquired fluorescence images with each mask (Fig.~\ref{fig2}c-d). The fluorescence enhancement increased with feature dimension, reaching a maximum of 52-fold at $\mathcal{D}=128$. The enhancement began to plateau beyond $\mathcal{D}=16$, where an enhancement of more than 49-fold, corresponding to over 94\% of the maximum, was achieved. At $\mathcal{D}=8$, the enhancement exceeded 40-fold, corresponding to approximately 80\% of the maximum. Thus, considering the diminishing improvement at larger feature dimensions, $\mathcal{D}\geq 8$ provides a favorable balance between fluorescence enhancement and computational complexity.

We next evaluated the convolutional kernel size $k$. The kernel size determines the spatial extent over which neighboring features are jointly processed. Because the correction masks are defined in the Fourier domain, the kernel size determines the range of neighboring spatial-frequency components coupled by each convolutional operation. Smaller kernels preserve sensitivity to local variations while maintaining low computational complexity, whereas larger kernels incorporate information over a broader neighborhood and may reduce sensitivity to localized variations in the correction mask. We evaluated kernel sizes ranging from 1 to 13 using the same set of fluorescence measurements (Fig.~\ref{fig2}e-f). The fluorescence enhancement generally decreased as the kernel size increased. For kernel sizes below 9, the fluorescence intensity varied only moderately and remained substantially higher than that obtained with larger kernels. Notably, the best performance was obtained with a kernel size of 1, for which no spatial mixing between neighboring features occurs. This result suggests that, under our experimental conditions, the benefit of the untrained-network parameterization does not primarily arise from convolutional feature extraction across neighboring spatial-frequency components. Instead, it is consistent with implicit regularization arising from gradient-based optimization of the overparameterized network representation, together with the explicit spatial regularization imposed by the TV term \cite{Heckel2019-dd, Heckel2020-cs}.

Finally, we investigated how the TV regularization coefficient influences scattering-correction performance. The TV coefficient controls the strength of spatial regularization applied to the correction mask. Because the mask is defined in the Fourier domain, the TV term promotes smoothness between neighboring spatial-frequency components. Increasing the TV coefficient can suppress noise and pixel-scale fluctuations but may also remove meaningful fine-scale features and sharply varying structures. We evaluated DeepFOCUS over a broad range of TV coefficients from $1\times10^{-5}$ to $3\times10^{-3}$, spanning weak to strong spatial regularization (Fig.~\ref{fig2}g-h). Within the range examined, fluorescence enhancement increased as the TV coefficient decreased. Considering both scattering-correction performance and the need to maintain sufficient spatial regularization, TV coefficients ranging from $1\times10^{-5}$ to $3\times10^{-4}$ were used in subsequent experiments.

In summary, these measurements quantitatively establish how the sensing and network parameters influence DeepFOCUS performance and provide an empirical parameter regime for subsequent in vivo experiments. Although the absolute fluorescence intensities and enhancement ratios varied among independent experiments, the best-performing condition in each parameter sweep consistently produced at least a 35-fold fluorescence enhancement. Based on these measurements, we selected 5,000 fluorescence measurements, a feature dimension of 16, a kernel size of 1, and a TV coefficient of $3\times10^{-4}$ as the default parameters for subsequent experiments. Although these parameters were optimized ex vivo, previous work \cite{He2025-au} demonstrated that parameters established using ex vivo samples can provide effective guidance for in vivo scattering correction. Together, these quantitative measurements establish a practical operating regime that balances scattering-correction performance, sensing time, and computational complexity for DeepFOCUS.

\subsection{Large-FOV in vivo vascular imaging to 1.1 mm depth with DeepFOCUS}\label{subsec2.3}

\begin{figure}[h!]
\centering
\includegraphics[width=1\textwidth]{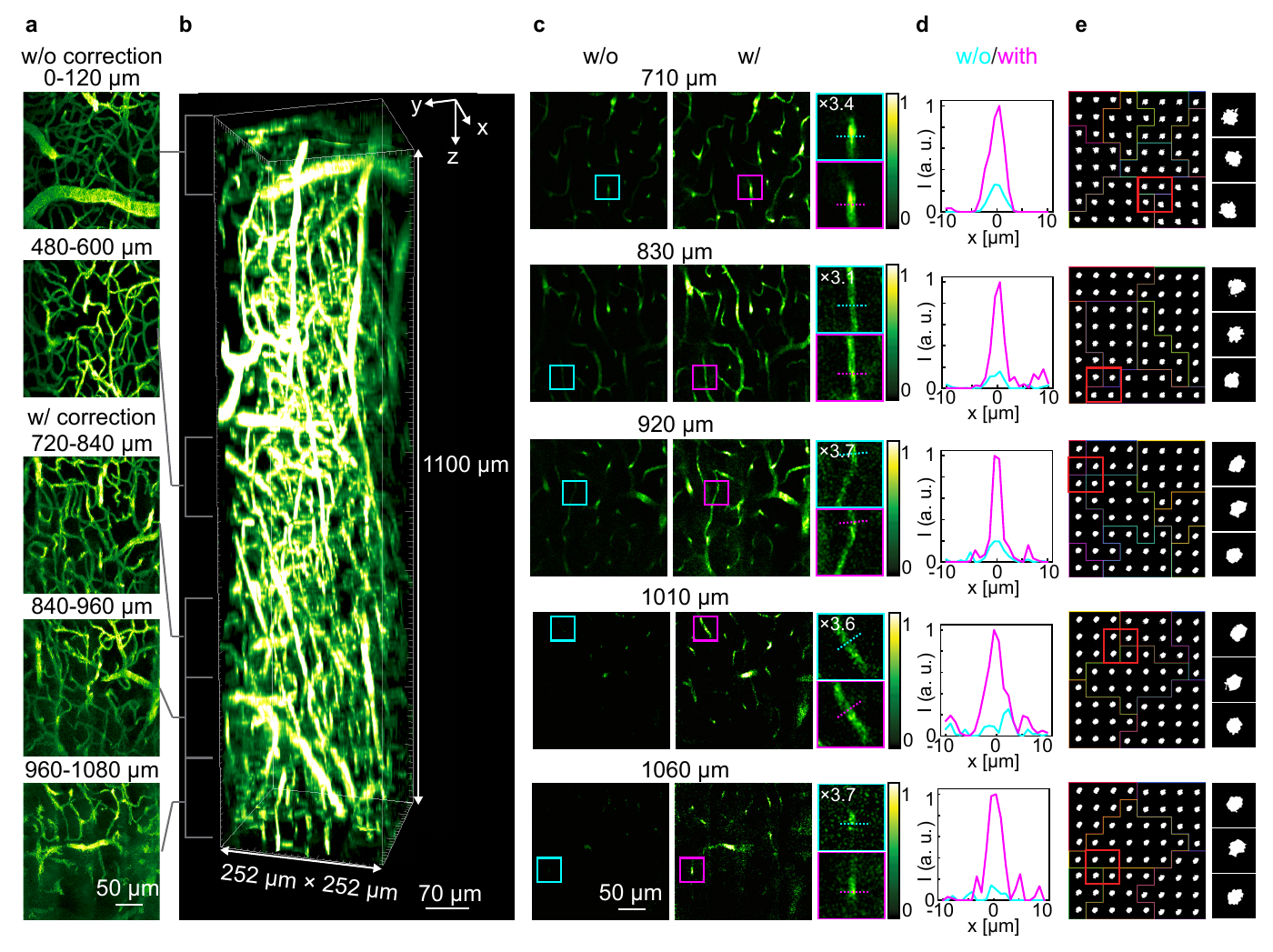}
\caption{\textbf{DeepFOCUS visualizes FITC-labeled blood vessels to 1.1 mm depth in the mouse brain in vivo.} Adult wild-type mice were imaged through a cranial window under anesthesia following intravenous injection of FITC. \textbf{a.} Maximum-intensity projections of selected depth ranges from the brain surface to 700 $\mu$m without correction and from 710 to 1100 $\mu$m with DeepFOCUS correction. \textbf{b.} Volumetric rendering of FITC-labeled blood vessels with scattering correction over a $252 \times 252 \times 1100~\mu\mathrm{m}^3$ imaging volume. \textbf{c.} Representative images acquired without correction (left), with correction (middle), and corresponding zoomed-in views (right) at 710, 830, 920, 1010, and 1060 $\mu$m depths. In the zoomed-in views, images without correction are digitally magnified by the factors indicated in each panel for visualization. \textbf{d.} Fluorescence intensity profiles along the dashed lines in the corresponding zoomed-in views in \textbf{c}, comparing images without and with correction. \textbf{e.} Subregion segmentation and representative correction masks corresponding to the red-boxed subregions at the depths shown in \textbf{c}.}\label{fig3}
\end{figure}

Next, we demonstrate DeepFOCUS for deep-brain imaging in vivo. We imaged FITC-labeled blood vessels over a $252 \times 252 \times 1100~\mu\mathrm{m}^3$ volume in the intact brain of a wild-type mouse through a cranial window. Details of animal preparation and imaging parameters are provided in \hyperref[Methods]{Methods} and Supplementary Table~\ref{TableS1}. Without scattering correction, blood vessels in the visual cortex could be imaged to a depth of 700 $\mu$m using a maximum laser power of 22.7 mW at the brain surface. Two representative maximum-intensity projections (MIPs; $z=0$--120 $\mu$m and 480--600 $\mu$m below the brain surface) reveal pial arteries at the surface, penetrating arterioles, and a dense capillary network deeper in the brain (Fig.~\ref{fig3}a-b).

We then applied multi-region scattering correction with DeepFOCUS from 700 to 1100 $\mu$m depth at 10-$\mu$m intervals along $z$. Each imaging plane was divided into 5--9 subregions to correct spatially varying scattering beyond a single isoplanatic patch. In representative images (Fig.~\ref{fig3}c), DeepFOCUS enhanced the fluorescence intensity by 3.4-fold at 710 $\mu$m, 3.1-fold at 830 $\mu$m, 3.7-fold at 920 $\mu$m, 3.6-fold at 1010 $\mu$m, and 3.7-fold at 1060 $\mu$m. DeepFOCUS also preserved fine spatial features at depth, resolving capillaries as small as 2.1 $\mu$m in diameter after correction at 920 $\mu$m (Fig.~\ref{fig3}d). The corresponding correction masks exhibited substantial spatial variation among subregions (Fig.~\ref{fig3}e), highlighting the importance of subregion-specific scattering correction. The corrected MIPs of deeper brain regions ($z=720$--840 $\mu$m, 840--960 $\mu$m, and 960--1080 $\mu$m; Fig.~\ref{fig3}a) exhibit contrast comparable to that of superficial regions imaged without correction, demonstrating substantial recovery of fluorescence signal at depth. Together, these results demonstrate that DeepFOCUS enables in vivo imaging of FITC-labeled blood vessels to a depth of 1100 $\mu$m (8.8 EALs, Fig.~\ref{figS2}, \cite{He2025-au}) in the intact mouse brain using 1035-nm excitation, over a $252\times252~\mu\mathrm{m}^2$ FOV while resolving capillaries as small as 2.1 $\mu$m in diameter. This imaging depth extends beyond the conventional scattering-limited regime of two-photon microscopy and reaches a depth range more commonly accessed using three-photon microscopy 

\subsection{DeepFOCUS enables high-resolution in vivo imaging to 1.2 mm depth in the hippocampus}\label{subsec2.4}

\begin{figure}[h!]%
\centering
\includegraphics[width=1\textwidth]{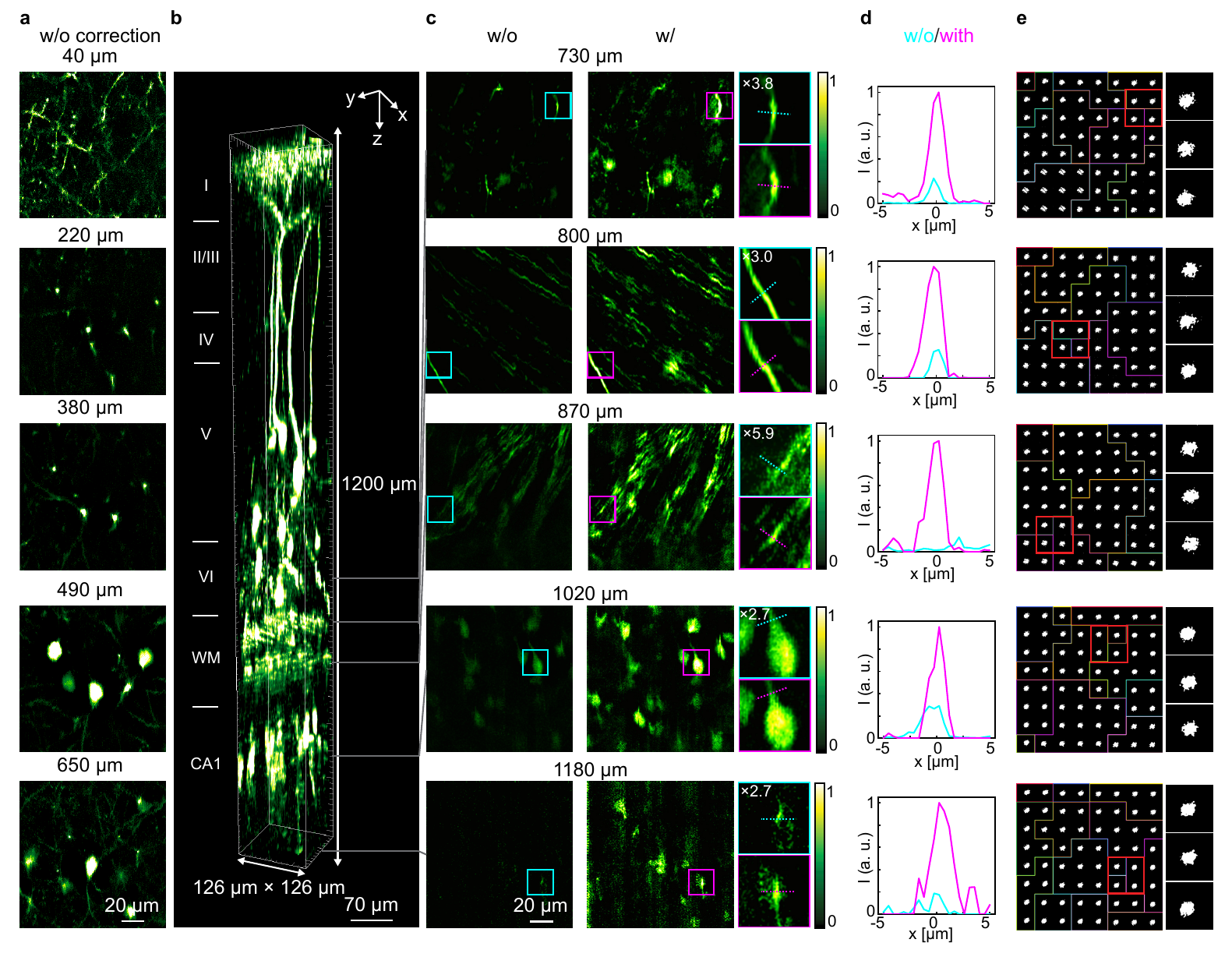}
\caption{\textbf{DeepFOCUS visualizes CA1 neurons to 1200 $\mu$m depth in the hippocampus in vivo.} An adult transgenic Thy1-YFP-H mouse was imaged through a cranial window under anesthesia. \textbf{a.} Representative images acquired without correction at depths from 40 to 650 $\mu$m. \textbf{b.} Volumetric view of YFP-labeled neurons with scattering correction over a $126 \times 126 \times 1200~\mu\mathrm{m}^3$ imaging volume. \textbf{c.} Representative images acquired without correction (left), with correction (middle), and corresponding zoomed-in views (right) at depths of 730, 800, 870, 1020, and 1180 $\mu$m. In the zoomed-in views, the intensities of the images without correction are digitally scaled by the factors indicated in each panel for visualization. \textbf{d.} Fluorescence intensity profiles along the dashed lines in the corresponding zoomed-in views in \textbf{c}, comparing images without and with correction. \textbf{e.} Subregions and representative correction masks at the corresponding depths shown in \textbf{c}. The subregions corresponding to the displayed masks are indicated by red boxes.}\label{fig4}
\end{figure}

Finally, we applied DeepFOCUS to perform continuous in vivo imaging from the brain surface through the cortical layers and white matter to the CA1 region of the hippocampus in the intact brain of transgenic Thy1-YFP-H mice through a cranial window. Details of animal preparation and imaging parameters are provided in \hyperref[Methods]{Methods}. The excitation power was increased with imaging depth, reaching a maximum of 100 mW at the sample surface. Without correction, we captured distinct anatomical structures characteristic of different depths down to 700 $\mu$m, including tuft dendrites near the cortical surface (representative image at 40 $\mu$m), apical dendrites within the superficial and middle cortical layers (220 and 380 $\mu$m), and neuronal somata in the deeper cortex (490 and 650 $\mu$m; Fig.~\ref{fig4}a). Beyond this depth, imaging reached the bottom of the cortex and entered the highly scattering white matter, where dense myelinated axon bundles substantially increased tissue scattering and degraded the fluorescence signal.

We then applied scattering correction from 700 to 1200 $\mu$m depth at 10-$\mu$m intervals along $z$. For each subregion, a correction mask was generated using 5,000 random patterns (40\% sparsity, $8\times8$-pixel DMD superpixels, and $100\times100$ superpixels per mask) and binarized using a DMD output-to-input power ratio of 34\%. With correction, we successfully visualized axons in layer 6 of the visual cortex at 730 $\mu$m depth and within the underlying white matter (Fig.~\ref{fig4}c). Notably, axons in the white matter formed two distinct layers at approximately 800 and 870 $\mu$m depth with different predominant orientations (Fig.~\ref{fig4}c). DeepFOCUS substantially enhanced the fluorescence signal, with increases of up to 5.9-fold in the white matter (Fig.~\ref{fig4}c). Below the white matter, imaging reached the hippocampus, a subcortical region critically involved in learning and memory. We clearly identified pyramidal neurons in the CA1 region at 1020 $\mu$m depth and resolved their dendritic processes extending to 1180 $\mu$m depth (Fig.~\ref{fig4}c). Zoomed-in views and fluorescence intensity profiles further show that DeepFOCUS resolves axons and dendrites with diameters of approximately 1.7--2.2 $\mu$m across layer 6, the white matter, and the CA1 region (Fig.~\ref{fig4}d). To correct strongly spatially varying scattering in the hippocampus, 13 distinct correction masks were generated over a $126\times126~\mu\mathrm{m}^2$ FOV at 1180 $\mu$m depth (Fig.~\ref{fig4}e). The measurement time was approximately 9 s per correction mask, and the computation time was approximately 6 s per mask, resulting in a total correction time of 191.1 s for all 13 subregions. Additional details on the correction parameters and videos of the image stacks are provided in \hyperref[Methods]{Methods} and Supplementary Information. In summary, DeepFOCUS enables in vivo two-photon imaging through the intact mouse brain to a depth of 1200 $\mu$m (9.4 EALs, Fig.~\ref{figS2}, \cite{He2025-au}), while resolving individual axons and dendrites within the white matter and hippocampal CA1 region. By overcoming the scattering-imposed depth limitation encountered by conventional two-photon microscopy, DeepFOCUS enables high-resolution, large-FOV imaging beyond 1 mm depth and provides direct optical access to the hippocampus in the intact mouse brain.

\section{Discussion}\label{sec12}

We have demonstrated that DeepFOCUS enables real-time correction of strong and spatially varying tissue scattering using intensity modulation computed by a self-supervised deep-learning framework. Unlike image-guided deep-learning approaches, DeepFOCUS does not rely on spatial information preserved by ballistic photons, but instead uses fluorescence generated by the interference of multiply scattered excitation light as feedback. Moreover, rather than applying CNNs in the image domain, DeepFOCUS parameterizes the correction mask in the Fourier domain, using the untrained network representation as an implicit ``spatial-frequency prior". Together, these advances enable DeepFOCUS to operate under substantially stronger scattering and at greater depths than previous scattering-correction approaches for two-photon microscopy. With 1035-nm excitation, DeepFOCUS enabled in vivo imaging of both FITC-labeled blood vessels and YFP-labeled neurons beyond 1 mm depth in the intact mouse brain, which was not achievable using this wavelength previously. Importantly, imaging beyond 1 mm was achieved while retaining micron-scale structural information and a FOV exceeding 100 $\mu$m through multi-region scattering correction. DeepFOCUS therefore provides optical access to deep-brain structures that are otherwise obscured by tissue scattering.

Several opportunities remain to further improve DeepFOCUS. The maximum depth demonstrated here is 1.2 mm (9.4 EALs), approaching but not yet reaching the approximately 10-EAL diffusion regime. Improving fluorescence collection during sensing and developing computational methods that are more robust to extremely low-SNR measurements could further extend the imaging depth. In addition, although sensing and computation are sufficiently rapid for the experiments demonstrated here, the overall correction time could be further reduced through faster sensing, reduced initialization and data-loading overhead, and more efficient computation.

In summary, DeepFOCUS is an intensity-based, self-supervised scattering-correction framework that extends two-photon microscopy beyond its conventional scattering-limited regime and enables in vivo imaging beyond 1 mm depth in the intact mouse brain. While the maximum optical imaging depth is still set by three-photon microscopy, DeepFOCUS is complementary rather than a replacement for three-photon microscopy. Importantly, DeepFOCUS provides a practical approach for laboratories to upgrade existing two-photon microscopes for noninvasive imaging of cortical--hippocampal projections and other deep-brain structures.

\section*{Methods}\label{Methods}

\textbf{Animals} 

All experimental and surgical protocols were approved by the University of California, Davis, Institutional Animal Care and Use Committee. One male/female wildtype mouse with age of weeks is used for imaging blood vessels. One male Thy1-YFP-H transgenic mouse with age of 11 weeks was used for imaging neurons. Both mice are from the Jackson Laboratories. Details on animal preparations are described below.  

\noindent\textbf{Optical setup}

The optical setup for DeepFOCUS was based on our previously described system \cite{He2025-au}. The system uses a femtosecond pulsed laser with a wavelength of 1035 nm and a repetition rate of 1 MHz (Monaco 1035-40-40 LX, Coherent). A polarizing beam splitter cube (PBS123, Thorlabs) and a half-wave plate (WPHSM05-1310, Thorlabs) mounted on a motorized rotation mount (K10CR2, Thorlabs) are used to adjust the laser power delivered to the downstream optics. Two one-axis galvanometric mirrors (QS7X-AG and QS7Y-AG, Thorlabs) are placed at a Fourier plane to scan the laser beam in two dimensions and are relayed by a 1:1 4-$f$ system (AC508-250-C-ML and ACT508-250-C-ML, Thorlabs). The laser beam is then expanded using a 4-$f$ system (LA1401-B and AC508-100-C-ML, Thorlabs) and pre-dispersed by a ruled diffraction grating (GR13-0310, 300 lines/mm, 1000-nm blaze, Thorlabs) to compensate for dispersion introduced by the DMD (DLP650LNIR, 1280$\times$800 pixels, maximum pattern rate of 12.5 kHz, VIALUX). The DMD is relayed by a 4-$f$ system consisting of two AC508-200-C-ML lenses (Thorlabs). The dispersion-compensation optics are omitted from the simplified optical schematic in Fig.~\ref{fig1}a. The beam reflected from the DMD is relayed to the back aperture of the objective lens through two additional 4-$f$ systems (AC508-150-C-ML and ACT508-300-C-ML; two ACT508-200-B-ML lenses, Thorlabs). A dichroic mirror (DMSP680B, Thorlabs) reflects the excitation beam toward the back aperture of the objective lens (XLUMPlanFL N, NA 1.0, $\times$20, Olympus), which is mounted on a motorized $z$-stage (V-308, Physik Instrumente) for axial scanning. In the emission path, a short-pass filter (ET750sp-2p8, Chroma) blocks residual excitation light. Fluorescence emission is spectrally filtered using a bandpass filter (AT535/40m, Chroma, for Fig.~\ref{fig1}d and Figs.~\ref{fig3}--\ref{fig4}; AT635/60m, Chroma, for Fig.~\ref{fig1}c and Fig.~\ref{fig2}) and detected by a photomultiplier tube (H15460, Hamamatsu). The system is controlled using MATLAB on a workstation (Precision 7960, Dell), with a data-acquisition card (PCIe-6363, X Series DAQ, National Instruments) providing analog and digital signal input/output.

\noindent\textbf{Typical parameters for scattering correction}

The specific parameters for all correction results are listed in Supplementary Table \ref{TableS1}--\ref{TableS2}.

\textbf{\textit{Data acquisition.}} Data acquisition was performed as previously described \cite{He2025-au} with modifications. During sensing, DeepFOCUS records the fluorescence intensity while sequentially projecting random binary patterns on the DMD. Each pattern consists of $100 \times 100$ superpixels, with each superpixel comprising $8 \times 8$ DMD pixels. The random binary patterns are generated from a uniform discrete distribution, with 40\% of the superpixels turned on. A total of 5,000 random patterns are used in the experiments shown in Fig.~\ref{fig1} and Figs.~\ref{fig3}--\ref{fig4}, and the patterns are preloaded onto the DMD before acquisition. The number of patterns used in Fig.~\ref{fig2} is specified in the main text. The projection time for each random pattern ranges from 0.5 ms (2 kHz) to 2 ms (0.5 kHz), with the specific settings listed in Supplementary Tables~\ref{TableS1}--\ref{TableS2}. Within each projection cycle, 0.1 ms is allocated for DMD pattern switching while the excitation laser is turned off, consistent with the maximum DMD switching rate of 12.5 kHz.

For point-scanning images acquired without correction or with global correction (Fig.~\ref{fig2}), either a blank mask or a single correction mask is displayed on the DMD while the galvanometric mirrors scan the FOV at 4 kHz. Each $z$-plane contains $288\times288$ pixels at 0.44 $\mu$m/pixel, with a 10-$\mu$m axial step. For subregion-specific correction (Figs.~\ref{fig3}--\ref{fig4}), DMD projection is synchronized with galvo scanning so that the appropriate mask is displayed for each subregion. The total exposure time is kept identical for corrected and uncorrected acquisitions.

\textbf{\textit{Content-aware subregions.}}
DeepFOCUS determines subregions following our previously described approach \cite{He2025-au}. The FOV is first divided into $8\times8$ patches, and the local intensity maximum within each patch is selected as the target. If a target is too dim or lies within a minimum separation of a neighboring target (30 $\mu$m for neurons and 60 $\mu$m for blood vessels), adjacent patches are merged into a content-aware subregion. The brightest peak within the merged region is then selected as the target for scattering correction.

\textbf{\textit{Computing correction masks.}} The correction masks were computed using a custom lightweight CNN implemented in Python on a GPU-equipped workstation (Precision 7960, Dell; GPU: NVIDIA RTX 4000 ADA). For each mask, the algorithm received an ($N \times 1$) vector of fluorescence measurements acquired under random-pattern modulation and the corresponding ($N \times 100 \times 100$) array of binary random patterns, where $N$ denotes the number of measurements. The network generated a grayscale correction mask comprising ($100 \times 100$) superpixels, which was subsequently binarized to achieve a DMD output-to-input power ratio of 30–40\% for Figs.~\ref{fig1}d, \ref{fig3}--\ref{fig4}, and 5–12\% for Figs. \ref{fig1}c and \ref{fig2}. The computation time per correction mask was 5–7 s for Fig. ~\ref{fig1}d, ~\ref{fig3}, \ref{fig4} and 11–14 s for Fig.~\ref {fig1}c and Fig.~\ref{fig2}.  The longer computation time in the latter experiments primarily resulted from including Python-environment initialization in the reported time for each single-mask calculation.

\noindent\textbf{Digital image processing}

Digital image processing was performed using MATLAB, ImageJ, and Imaris Viewer. Raw images were denoised using notch and median filters, followed by background subtraction using unsharp masking. For images acquired with subregion correction, patch boundaries were refined by locally compensating for intensity nonuniformities near the margins, typically within a 1–3-pixel-wide border. Volumetric renderings were generated in Imaris Viewer using gamma values of 0.5–1.0 and saturation values of 0.7–0.8 for visualization. Two-dimensional images were displayed in ImageJ using the “Green Hot” colormap, with contrast adjusted for visualization.

\noindent\textbf{Bead sample preparation}

Fluorescent bead samples were prepared as previously described \cite{He2025-au}. Red fluorescent beads (R700, Thermo Fisher Scientific, MA) were mixed with PDMS (Sylgard 184, Dow Inc., MI) and covered with a coverslip. Air bubbles were removed using a vacuum desiccator, and the sample was cured at 100 $^\circ$C for 35 min. After curing, a dissected mouse skull approximately 3 mm in diameter was attached to the top of the coverslip using superglue (Gorilla).

\noindent\textbf{Mouse craniotomy for imaging through a cranial window}

A survival craniotomy was performed to enable imaging of neurons or blood vessels (Figs.~\ref{fig1}d, \ref{fig3}, and \ref{fig4}) as previously described \cite{He2025-au}. During surgery, mice were anesthetized with 2--3\% isoflurane delivered through a vaporizer. The head was stabilized using ear bars and a tooth bar on a stereotactic frame (51730, Stoelting). Toe-pinch reflexes were monitored throughout the procedure to assess anesthetic depth. Body temperature was maintained at 37 $^\circ$C using a heating pad (53800, Stoelting) and monitored with a rectal probe. Ophthalmic ointment was applied to protect the eyes. Fur over the surgical area was removed, and the scalp was sterilized with alternating applications of chlorhexidine and ethanol. The scalp was then excised to expose the skull, and the periosteum was removed with a scalpel blade. A 3-mm-diameter craniotomy was marked using a biopsy punch, and the skull was gradually thinned and removed using a micromotor drill (51449, Stoelting), while keeping the dura intact.

The cranial window was prepared before surgery by bonding two coverslips (3 and 5 mm in diameter) with ultraviolet-cured optical adhesive (Norland NOA 65). The window was placed over the craniotomy and secured with dental cement (Parkell C\&B Metabond), which was also used to seal the wound margins. After the cement hardened, a stainless-steel headplate was attached using the same dental cement. Imaging was performed at least one week after surgery. During imaging sessions, mice were anesthetized with 1.5--2\% isoflurane and head-fixed.

For vascular imaging (Fig.~\ref{fig3}), FITC-dextran (2.5\%, 1 g/40 mL in saline; 0.2 mL) was administered by retro-orbital injection immediately before imaging. The mouse was euthanized immediately after the imaging session. For neuronal imaging (Fig.~\ref{fig4}), no additional fluorescent labeling was required because a transgenic Thy1-YFP-H mouse was used. After imaging, the mouse was returned to the vivarium following recovery from anesthesia.

\backmatter

\bmhead{Author contributions}
Y.X. led the project and mentored R.H. and Y.L. Y.X. designed and built the microscopy system and performed the mouse preparation. Y.L., R.H., and Y.X. developed the data acquisition and processing code. Y.L. performed the experiments and acquired the experimental data. Y.L. and R.H. processed and analyzed the experimental data. All authors contributed to writing and revising the manuscript.

\bmhead{Acknowledgements}
We thank Meng Wang at UC Davis for providing the FITC solution used for vascular imaging. 

\bmhead{Funding}
Research reported in this publication was primarily supported by the National Institute of General Medical Sciences of the National Institutes of Health under Award Number R35GM155193 and by the National Science Foundation CAREER Award 2443604 to Yi Xue. This work was also supported by startup funds provided to Yi Xue by the Department of Biomedical Engineering at the University of California, Davis.

\bmhead{Data Availability}
The main data supporting the findings of this study are available within the paper and its Supplementary Information files.

\bmhead{Code Availability}
The code used in this study is available from the corresponding author upon reasonable request.

\section*{Declarations}
The authors declare no competing interests.

\bibliography{sn-bibliography}% common bib file

@ARTICLE{Rodriguez2021-sy,
  title    = "An adaptive optics module for deep tissue multiphoton imaging in
              vivo",
  author   = "Rodríguez, Cristina and Chen, Anderson and Rivera, José A and
              Mohr, Manuel A and Liang, Yajie and Natan, Ryan G and Sun, Wenzhi
              and Milkie, Daniel E and Bifano, Thomas G and Chen, Xiaoke and Ji,
              Na",
  journal  = "Nat. Methods",
  volume   =  18,
  number   =  10,
  pages    = "1259--1264",
  month    =  oct,
  year     =  2021,
  language = "en"
}

@ARTICLE{He2025-au,
  title         = "Compressive Fourier-Domain Intensity Coupling ({C}-{FOCUS})
                   enables near-millimeter deep imaging in the intact mouse
                   brain in vivo",
  author        = "He, Renzhi and Li, Yucheng and Urbina, Brianna and Wan,
                   Jiandi and Xue, Yi",
  journal       = "arXiv [physics.optics]",
  month         =  may,
  year          =  2025,
  archivePrefix = "arXiv",
  primaryClass  = "physics.optics"
}

@ARTICLE{Kang2024-vy,
  title     = "Coordinate-based neural representations for computational
               adaptive optics in widefield microscopy",
  author    = "Kang, Iksung and Zhang, Qinrong and Yu, Stella X and Ji, Na",
  journal   = "Nat. Mach. Intell.",
  publisher = "Springer Science and Business Media LLC",
  volume    =  6,
  number    =  6,
  pages     = "714--725",
  month     =  jun,
  year      =  2024,
  language  = "en"
}

@ARTICLE{Kang2026-ay,
  title     = "Adaptive optical correction for in vivo two-photon fluorescence
               microscopy with neural fields",
  author    = "Kang, Iksung and Kim, Hyeonggeon and Natan, Ryan and Zhang,
               Qinrong and Yu, Stella X and Ji, Na",
  journal   = "Nat. Methods",
  publisher = "Springer Science and Business Media LLC",
  pages     = "1--10",
  month     =  apr,
  year      =  2026,
  language  = "en"
}

@ARTICLE{Chen2021-kq,
  title     = "High-resolution two-photon transcranial imaging of brain using
               direct wavefront sensing",
  author    = "Chen, Congping and Qin, Zhongya and He, Sicong and Liu, Shaojun
               and Lau, Shun-Fat and Wu, Wanjie and Zhu, Dan and Ip, Nancy Y and
               Qu, Jianan Y",
  journal   = "Photonics Res.",
  publisher = "Optica Publishing Group",
  volume    =  9,
  number    =  6,
  pages     =  1144,
  month     =  jun,
  year      =  2021,
  language  = "en"
}

@ARTICLE{Papadopoulos2016-sy,
  title     = "Scattering compensation by focus scanning holographic aberration
               probing ({F}-{SHARP})",
  author    = "Papadopoulos, Ioannis N and Jouhanneau, Jean-Sébastien and
               Poulet, James F A and Judkewitz, Benjamin",
  journal   = "Nat. Photonics",
  publisher = "Nature Research",
  volume    =  11,
  number    =  2,
  pages     = "116--123",
  month     =  dec,
  year      =  2016,
  language  = "en"
}

@ARTICLE{Zepeda2025-cm,
  title     = "Scattering correction through Fourier-domain intensity coupling
               in two-photon microscopy ({2P}-{FOCUS})",
  author    = "Zepeda, Daniel and Li, Yucheng and Xue, Yi",
  journal   = "Photonics Res.",
  publisher = "Optica Publishing Group",
  volume    =  13,
  number    =  4,
  pages     =  845,
  month     =  apr,
  year      =  2025,
  language  = "en"
}

@ARTICLE{Kondo2017-lh,
  title     = "Two-photon calcium imaging of the medial prefrontal cortex and
               hippocampus without cortical invasion",
  author    = "Kondo, Masashi and Kobayashi, Kenta and Ohkura, Masamichi and
               Nakai, Junichi and Matsuzaki, Masanori",
  journal   = "Elife",
  publisher = "eLife Sciences Publications Limited",
  volume    =  6,
  month     =  sep,
  year      =  2017,
  language  = "en"
}

@ARTICLE{Mittmann2011-jo,
  title     = "Two-photon calcium imaging of evoked activity from {L5}
               somatosensory neurons in vivo",
  author    = "Mittmann, Wolfgang and Wallace, Damian J and Czubayko, Uwe and
               Herb, Jan T and Schaefer, Andreas T and Looger, Loren L and Denk,
               Winfried and Kerr, Jason N D",
  journal   = "Nat. Neurosci.",
  publisher = "Springer Science and Business Media LLC",
  volume    =  14,
  number    =  8,
  pages     = "1089--1093",
  month     =  jul,
  year      =  2011,
  language  = "en"
}

@ARTICLE{Wang2014-wo,
  title     = "Rapid adaptive optical recovery of optimal resolution over large
               volumes",
  author    = "Wang, Kai and Milkie, Daniel E and Saxena, Ankur and Engerer,
               Peter and Misgeld, Thomas and Bronner, Marianne E and Mumm, Jeff
               and Betzig, Eric",
  journal   = "Nat. Methods",
  publisher = "Springer Science and Business Media LLC",
  volume    =  11,
  number    =  6,
  pages     = "625--628",
  month     =  jun,
  year      =  2014,
  language  = "en"
}

@ARTICLE{Ouzounov2017-fk,
  title    = "In vivo three-photon imaging of activity of {GCaMP6}-labeled
              neurons deep in intact mouse brain",
  author   = "Ouzounov, Dimitre G and Wang, Tianyu and Wang, Mengran and Feng,
              Danielle D and Horton, Nicholas G and Cruz-Hernández, Jean C and
              Cheng, Yu-Ting and Reimer, Jacob and Tolias, Andreas S and
              Nishimura, Nozomi and Xu, Chris",
  journal  = "Nat. Methods",
  volume   =  14,
  number   =  4,
  pages    = "388--390",
  month    =  apr,
  year     =  2017,
  language = "en"
}

@ARTICLE{Helmchen2005-lg,
  title    = "Deep tissue two-photon microscopy",
  author   = "Helmchen, Fritjof and Denk, Winfried",
  journal  = "Nat. Methods",
  volume   =  2,
  number   =  12,
  pages    = "932--940",
  month    =  dec,
  year     =  2005,
  language = "en"
}

@ARTICLE{Horton2013-id,
  title    = "In vivo three-photon microscopy of subcortical structures within
              an intact mouse brain",
  author   = "Horton, Nicholas G and Wang, Ke and Kobat, Demirhan and Clark,
              Catharine G and Wise, Frank W and Schaffer, Chris B and Xu, Chris",
  journal  = "Nat. Photonics",
  volume   =  7,
  number   =  3,
  pages    = "205--209",
  month    =  mar,
  year     =  2013,
  language = "en"
}

@ARTICLE{Streich2021-wf,
  title    = "High-resolution structural and functional deep brain imaging using
              adaptive optics three-photon microscopy",
  author   = "Streich, Lina and Boffi, Juan Carlos and Wang, Ling and Alhalaseh,
              Khaleel and Barbieri, Matteo and Rehm, Ronja and Deivasigamani,
              Senthilkumar and Gross, Cornelius T and Agarwal, Amit and
              Prevedel, Robert",
  journal  = "Nat. Methods",
  volume   =  18,
  number   =  10,
  pages    = "1253--1258",
  month    =  oct,
  year     =  2021,
  language = "en"
}

@ARTICLE{Drew2010-bp,
  title    = "Chronic optical access through a polished and reinforced thinned
              skull",
  author   = "Drew, Patrick J and Shih, Andy Y and Driscoll, Jonathan D and
              Knutsen, Per Magne and Blinder, Pablo and Davalos, Dimitrios and
              Akassoglou, Katerina and Tsai, Philbert S and Kleinfeld, David",
  journal  = "Nat. Methods",
  volume   =  7,
  number   =  12,
  pages    = "981--984",
  month    =  dec,
  year     =  2010,
  language = "en"
}

@ARTICLE{Aviles-Espinosa2011-vt,
  title     = "Measurement and correction of in vivo sample aberrations
               employing a nonlinear guide-star in two-photon excited
               fluorescence microscopy",
  author    = "Aviles-Espinosa, Rodrigo and Andilla, Jordi and Porcar-Guezenec,
               Rafael and Olarte, Omar E and Nieto, Marta and Levecq, Xavier and
               Artigas, David and Loza-Alvarez, Pablo",
  journal   = "Biomed. Opt. Express",
  publisher = "Optica Publishing Group",
  volume    =  2,
  number    =  11,
  pages     = "3135--3149",
  month     =  nov,
  year      =  2011,
  language  = "en"
}

@ARTICLE{Albert2000-oj,
  title     = "Smart microscope: an adaptive optics learning system for
               aberration correction in multiphoton confocal microscopy",
  author    = "Albert, O and Sherman, L and Mourou, G and Norris, T B and
               Vdovin, G",
  journal   = "Opt. Lett.",
  publisher = "Optica Publishing Group",
  volume    =  25,
  number    =  1,
  pages     = "52--54",
  month     =  jan,
  year      =  2000,
  language  = "en"
}

@ARTICLE{Rueckel2006-lb,
  title    = "Adaptive wavefront correction in two-photon microscopy using
              coherence-gated wavefront sensing",
  author   = "Rueckel, Markus and Mack-Bucher, Julia A and Denk, Winfried",
  journal  = "Proc. Natl. Acad. Sci. U. S. A.",
  volume   =  103,
  number   =  46,
  pages    = "17137--17142",
  month    =  nov,
  year     =  2006,
  language = "en"
}

@ARTICLE{Debarre2009-ax,
  title    = "Image-based adaptive optics for two-photon microscopy",
  author   = "Débarre, Delphine and Botcherby, Edward J and Watanabe, Tomoko and
              Srinivas, Shankar and Booth, Martin J and Wilson, Tony",
  journal  = "Opt. Lett.",
  volume   =  34,
  number   =  16,
  pages    = "2495--2497",
  month    =  aug,
  year     =  2009,
  language = "en"
}

@ARTICLE{Cha2010-qe,
  title    = "Shack-Hartmann wavefront-sensor-based adaptive optics system for
              multiphoton microscopy",
  author   = "Cha, Jae Won and Ballesta, Jerome and So, Peter T C",
  journal  = "J. Biomed. Opt.",
  volume   =  15,
  number   =  4,
  pages    =  046022,
  month    =  jul,
  year     =  2010,
  language = "en"
}

@ARTICLE{Marsh2003-rh,
  title     = "Practical implementation of adaptive optics in multiphoton
               microscopy",
  author    = "Marsh, P and Burns, D and Girkin, J",
  journal   = "Opt. Express",
  publisher = "Optica Publishing Group",
  volume    =  11,
  number    =  10,
  pages     = "1123--1130",
  month     =  may,
  year      =  2003,
  language  = "en"
}

@ARTICLE{May2021-er,
  title    = "Fast holographic scattering compensation for deep tissue
              biological imaging",
  author   = "May, Molly A and Barré, Nicolas and Kummer, Kai K and Kress,
              Michaela and Ritsch-Marte, Monika and Jesacher, Alexander",
  journal  = "Nat. Commun.",
  volume   =  12,
  number   =  1,
  pages    =  4340,
  month    =  jul,
  year     =  2021,
  language = "en"
}

@ARTICLE{Tang2012-bl,
  title    = "Superpenetration optical microscopy by iterative multiphoton
              adaptive compensation technique",
  author   = "Tang, Jianyong and Germain, Ronald N and Cui, Meng",
  journal  = "Proc. Natl. Acad. Sci. U. S. A.",
  volume   =  109,
  number   =  22,
  pages    = "8434--8439",
  month    =  may,
  year     =  2012,
  language = "en"
}

@ARTICLE{Qin2022-ns,
  title    = "Deep tissue multi-photon imaging using adaptive optics with direct
              focus sensing and shaping",
  author   = "Qin, Zhongya and She, Zhentao and Chen, Congping and Wu, Wanjie
              and Lau, Jackie K Y and Ip, Nancy Y and Qu, Jianan Y",
  journal  = "Nat. Biotechnol.",
  volume   =  40,
  number   =  11,
  pages    = "1663--1671",
  month    =  nov,
  year     =  2022,
  language = "en"
}

@ARTICLE{Liu2019-ae,
  title     = "Direct wavefront sensing enables functional imaging of
               infragranular axons and spines",
  author    = "Liu, Rui and Li, Zengyi and Marvin, Jonathan S and Kleinfeld,
               David",
  journal   = "Nat. Methods",
  publisher = "Springer Science and Business Media LLC",
  volume    =  16,
  number    =  7,
  pages     = "615--618",
  month     =  jul,
  year      =  2019,
  language  = "en"
}

@ARTICLE{Ji2017-vf,
  title    = "Adaptive optical fluorescence microscopy",
  author   = "Ji, Na",
  journal  = "Nat. Methods",
  volume   =  14,
  number   =  4,
  pages    = "374--380",
  month    =  mar,
  year     =  2017,
  language = "en"
}

@ARTICLE{Wang2015-fl,
  title    = "Direct wavefront sensing for high-resolution in vivo imaging in
              scattering tissue",
  author   = "Wang, Kai and Sun, Wenzhi and Richie, Christopher T and Harvey,
              Brandon K and Betzig, Eric and Ji, Na",
  journal  = "Nat. Commun.",
  volume   =  6,
  pages    =  7276,
  month    =  jun,
  year     =  2015,
  language = "en"
}

@ARTICLE{Takasaki2020-od,
  title     = "Superficial bound of the depth limit of two-photon imaging in
               mouse brain",
  author    = "Takasaki, Kevin and Abbasi-Asl, Reza and Waters, Jack",
  journal   = "eNeuro",
  publisher = "Society for Neuroscience",
  volume    =  7,
  number    =  1,
  pages     = "ENEURO.0255--19.2019",
  month     =  jan,
  year      =  2020,
  language  = "en"
}

@ARTICLE{Cheng2019-zy,
  title    = "Deep-brain 2-photon fluorescence microscopy in vivo excited at the
              1700 nm window",
  author   = "Cheng, Hui and Tong, Shen and Deng, Xiangquan and Liu, Hongji and
              Du, Yu and He, Chen and Qiu, Ping and Wang, Ke",
  journal  = "Opt. Lett.",
  volume   =  44,
  number   =  17,
  pages    = "4432--4435",
  month    =  sep,
  year     =  2019,
  language = "en"
}

@ARTICLE{Nishizaki2019-bx,
  title     = "Deep learning wavefront sensing",
  author    = "Nishizaki, Yohei and Valdivia, Matias and Horisaki, Ryoichi and
               Kitaguchi, Katsuhisa and Saito, Mamoru and Tanida, Jun and Vera,
               Esteban",
  journal   = "Opt. Express",
  publisher = "Optica Publishing Group",
  volume    =  27,
  number    =  1,
  pages     = "240--251",
  month     =  jan,
  year      =  2019,
  language  = "en"
}

@ARTICLE{Paine2018-oh,
  title     = "Machine learning for improved image-based wavefront sensing",
  author    = "Paine, Scott W and Fienup, James R",
  journal   = "Opt. Lett.",
  publisher = "Optica Publishing Group",
  volume    =  43,
  number    =  6,
  pages     = "1235--1238",
  month     =  mar,
  year      =  2018,
  language  = "en"
}

@ARTICLE{Wu2020-uu,
  title     = "Sub-millisecond phase retrieval for phase-diversity wavefront
               sensor",
  author    = "Wu, Yu and Guo, Youming and Bao, Hua and Rao, Changhui",
  journal   = "Sensors (Basel)",
  publisher = "MDPI AG",
  volume    =  20,
  number    =  17,
  pages     = "E4877",
  month     =  aug,
  year      =  2020,
  language  = "en"
}

@ARTICLE{Hu2023-jt,
  title     = "Universal adaptive optics for microscopy through embedded neural
               network control",
  author    = "Hu, Qi and Hailstone, Martin and Wang, Jingyu and Wincott,
               Matthew and Stoychev, Danail and Atilgan, Huriye and Gala, Dalia
               and Chaiamarit, Tai and Parton, Richard M and Antonello, Jacopo
               and Packer, Adam M and Davis, Ilan and Booth, Martin J",
  journal   = "Light Sci. Appl.",
  publisher = "Springer Science and Business Media LLC",
  volume    =  12,
  number    =  1,
  pages     =  270,
  month     =  nov,
  year      =  2023,
  language  = "en"
}

@ARTICLE{Kobat2011-ki,
  title    = "\textit{In vivo} two-photon microscopy to 1.6-mm depth in mouse
              cortex",
  author   = "Kobat, Demirhan and Horton, Nicholas G and Xu, Chris",
  journal  = "J. Biomed. Opt.",
  volume   =  16,
  number   =  10,
  pages    =  106014,
  month    =  oct,
  year     =  2011,
  language = "en"
}

@ARTICLE{Yildirim2019-wn,
  title     = "Functional imaging of visual cortical layers and subplate in
               awake mice with optimized three-photon microscopy",
  author    = "Yildirim, Murat and Sugihara, Hiroki and So, Peter T C and Sur,
               Mriganka",
  journal   = "Nat. Commun.",
  publisher = "Springer Science and Business Media LLC",
  volume    =  10,
  number    =  1,
  pages     =  177,
  month     =  jan,
  year      =  2019,
  language  = "en"
}

@ARTICLE{Xin2019-px,
  title     = "Object-independent image-based wavefront sensing approach using
               phase diversity images and deep learning",
  author    = "Xin, Qi and Ju, Guohao and Zhang, Chunyue and Xu, Shuyan",
  journal   = "Opt. Express",
  publisher = "Optica Publishing Group",
  volume    =  27,
  number    =  18,
  pages     = "26102--26119",
  month     =  sep,
  year      =  2019,
  language  = "en"
}

@ARTICLE{Saha2020-il,
  title     = "Practical sensorless aberration estimation for {3D} microscopy
               with deep learning",
  author    = "Saha, Debayan and Schmidt, Uwe and Zhang, Qinrong and Barbotin,
               Aurelien and Hu, Qi and Ji, Na and Booth, Martin J and Weigert,
               Martin and Myers, Eugene W",
  journal   = "Opt. Express",
  publisher = "Optica Publishing Group",
  volume    =  28,
  number    =  20,
  pages     = "29044--29053",
  month     =  sep,
  year      =  2020,
  language  = "en"
}

@ARTICLE{Feng2023-ln,
  title     = "{NeuWS}: Neural wavefront shaping for guidestar-free imaging
               through static and dynamic scattering media",
  author    = "Feng, Brandon Y and Guo, Haiyun and Xie, Mingyang and
               Boominathan, Vivek and Sharma, Manoj K and Veeraraghavan, Ashok
               and Metzler, Christopher A",
  journal   = "Sci. Adv.",
  publisher = "American Association for the Advancement of Science",
  volume    =  9,
  number    =  26,
  pages     = "eadg4671",
  month     =  jun,
  year      =  2023,
  language  = "en"
}

@article{Wang:20,
author = {Tianyu Wang and Chris Xu},
journal = {Optica},
number = {8},
pages = {947--960},
publisher = {Optica Publishing Group},
title = {Three-photon neuronal imaging in deep mouse brain},
volume = {7},
month = {Aug},
year = {2020},
url = {https://opg.optica.org/optica/abstract.cfm?URI=optica-7-8-947},
doi = {10.1364/OPTICA.395825},
}

@InProceedings{Ulyanov_2018_CVPR,
author = {Ulyanov, Dmitry and Vedaldi, Andrea and Lempitsky, Victor},
title = {Deep Image Prior},
booktitle = {Proceedings of the IEEE Conference on Computer Vision and Pattern Recognition (CVPR)},
month = {June},
year = {2018}
}

@InProceedings{Heckel2019-dd,
author = {Heckel, Reinhard and Hand, Paul},
title = {Deep Decoder: Concise Image Representations from Untrained Non-convolutional Networks},
booktitle = {International Conference on Learning Representations (ICLR)},
year = {2019}
}

@InProceedings{Heckel2020-cs,
author = {Heckel, Reinhard and Soltanolkotabi, Mahdi},
title = {Compressive Sensing with Un-trained Neural Networks: Gradient Descent Finds a Smooth Approximation},
booktitle = {Proceedings of the 37th International Conference on Machine Learning (ICML)},
pages = {4149--4158},
year = {2020}
}

@Article{Geiger2020-sc,
author = {Geiger, Mario and Jacot, Arthur and Spigler, Stefano and Gabriel, Franck and Sagun, Levent and d'Ascoli, St{\'e}phane and Biroli, Giulio and Hongler, Cl{\'e}ment and Wyart, Matthieu},
title = {Scaling Description of Generalization with Number of Parameters in Deep Learning},
journal = {Journal of Statistical Mechanics: Theory and Experiment},
volume = {2020},
number = {2},
pages = {023401},
year = {2020}
}

@InProceedings{Jacot2018-ntk,
author = {Jacot, Arthur and Gabriel, Franck and Hongler, Cl{\'e}ment},
title = {Neural Tangent Kernel: Convergence and Generalization in Neural Networks},
booktitle = {Advances in Neural Information Processing Systems (NeurIPS)},
pages = {8580--8589},
year = {2018}
}

\newpage

\begin{appendices}

\counterwithout{figure}{section}
\counterwithout{table}{section}

\setcounter{figure}{0}
\setcounter{table}{0}

\renewcommand{\thefigure}{S\arabic{figure}}
\renewcommand{\thetable}{S\arabic{table}}

\section{Supplementary Figure}
\begin{figure}[h!]%
\centering
\includegraphics[width=1\textwidth]{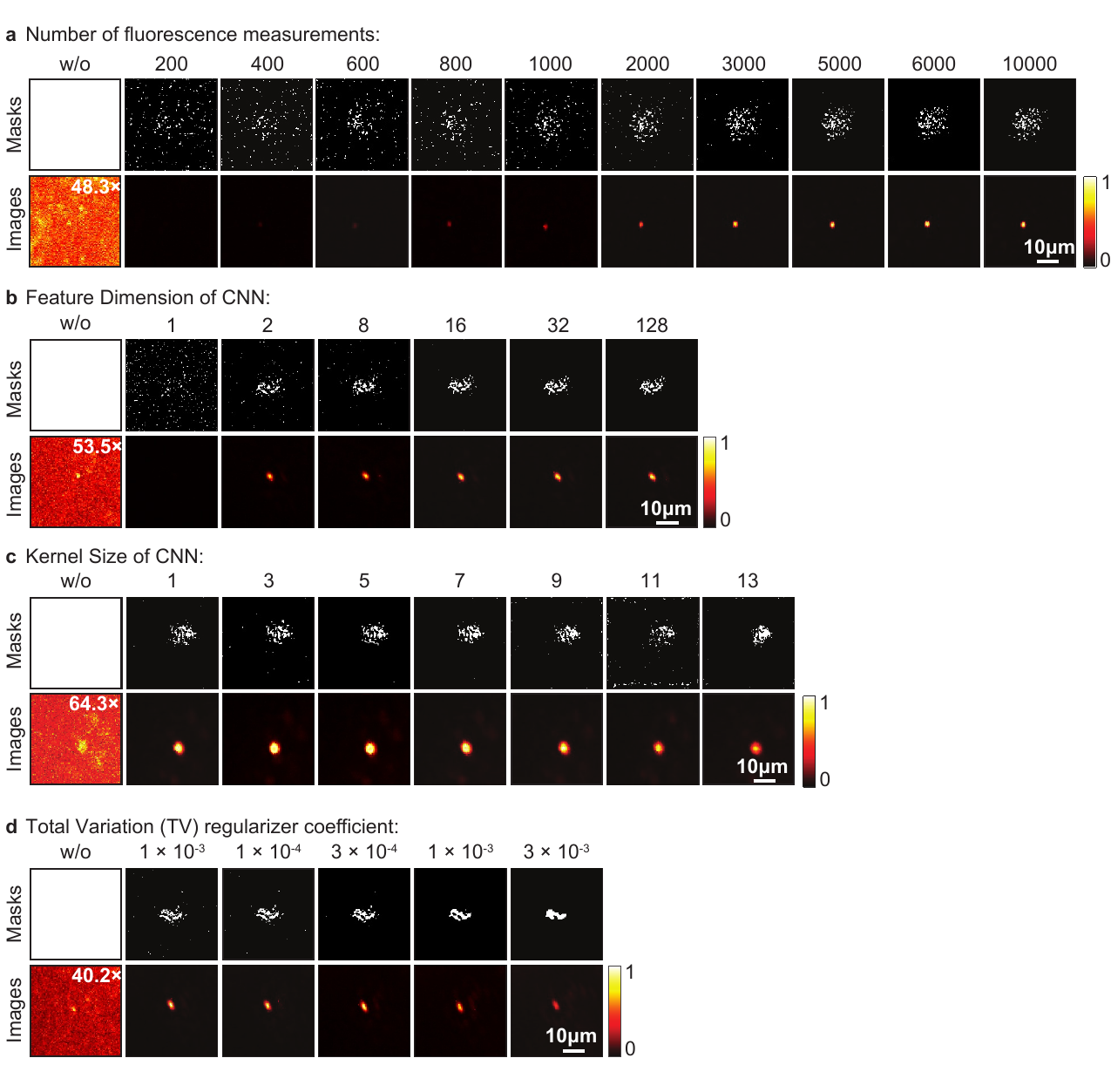}
\caption{\textbf{\boldmath Quantitative evaluation of DeepFOCUS by imaging red fluorescent beads through a 250 $\mu$m mouse skull ex vivo. }
\textbf{a} Fluorescence images of a bead acquired with an all-white mask (uncorrected; displayed at 48.3-fold magnification) and with correction masks computed from 200–10,000 measurements, together with the corresponding correction masks. 
\textbf{b} Fluorescence images of a bead acquired with an all-white mask (uncorrected; displayed at 53.5-fold magnification) and with correction masks generated using CNNs with feature dimensions ranging from 1 to 128, together with the corresponding correction masks. 
\textbf{c.} Fluorescence images of a bead acquired with an all-white mask (uncorrected; displayed at 64.3-fold magnification) and with correction masks generated using CNNs with kernel sizes ranging from 1 to 13, together with the corresponding correction masks. 
\textbf{d.} Fluorescence images of a bead acquired with an all-white mask (uncorrected; displayed at 40.2-fold magnification) and with correction masks generated using CNNs with TV regularization coefficients ranging from  $1 \times 10^{-5}$ to $3 \times 10^{-3}$, together with the corresponding correction masks. 
}\label{figS1}
\end{figure}

\begin{figure}[h!]%
\centering
\includegraphics[width=1\textwidth]{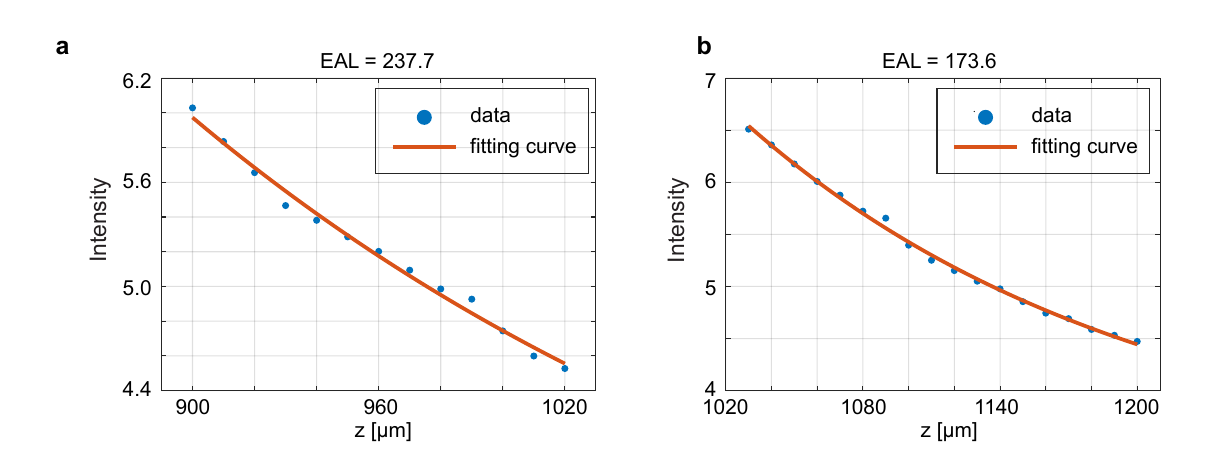}
\caption{\textbf{Experimentally measured effective attenuation lengths (EALs) in the mouse CA1 region in vivo.} EALs were obtained by fitting the estimated excitation intensity to an exponential decay, $I_z = I_0 e^{-z/\mathrm{EAL}}$, where $I_z$ and $I_0$ denote the estimated laser intensities (arbitrary units, a.u.) at depth $z$ and at the initial imaging plane, respectively. The excitation intensity was estimated from the two-photon fluorescence signal as $I=\sqrt{I_{\mathrm{fluo}}}$. \textbf{a,b.} In vivo EALs measured in the CA1 region of a Thy1-YFP-H mouse: \textbf{a,} 237.7 $\mu$m over 900--1020 $\mu$m depth; \textbf{b,} 173.6 $\mu$m over 1020--1200 $\mu$m depth. EALs from the brain surface to 900 $\mu$m depth were reported previously \cite{He2025-au} and were used in this study.}\label{figS2}
\end{figure}

\section{Supplementary Table}\label{TableS1}
\begin{sidewaystable}
\centering
\caption{CNN parameters, computation time and measurement time used for the scattering correction in Fig.~\ref{fig2}, \ref{figS1}.}
\label{TableS1}
\small
\setlength{\tabcolsep}{4pt}       % default is 6 pt
\renewcommand{\arraystretch}{0.85} % default is 1
\begin{tabular*}{0.85\textheight}{ccccccc}
\toprule
Fig. & Measurements & Kernel size & TV Coefficients & Feature Dimension & Computation Time & Measurement time\\
\midrule
A1a & 200 & 1  & $1 \times 10^{-4}$ & 128 & 12.4882  & 3.6372  \\
A1a & 400 & 1  & $1 \times 10^{-4}$ & 128 & 11.4742  & 4.3752  \\
A1a & 600 & 1  & $1 \times 10^{-4}$ & 128 & 12.2296  & 4.5419 \\
A1a & 800 & 1  & $1 \times 10^{-4}$ & 128 & 12.0083  & 4.8555 \\
A1a & 1000 & 1  & $1 \times 10^{-4}$ & 128 & 12.3435 & 5.8089  \\
A1a & 2000 & 1  & $1 \times 10^{-4}$ & 128 & 12.3281  & 7.6562  \\
A1a & 3000 & 1  & $1 \times 10^{-4}$ & 128 & 12.2549 & 9.7304 \\
A1a & 5000 & 1  & $1 \times 10^{-4}$ & 128 & 11.9917 & 14.6802 \\
A1a & 6000 & 1  & $1 \times 10^{-4}$ & 128 & 11.9151 & 16.3856 \\
A1a & 10000 & 1  & $1 \times 10^{-4}$ & 128 & 13.0124 & 26.4071 \\
A1b & 5000 & 1  & $3 \times 10^{-4}$ & 1 & 11.2393 & 14.2225 \\
A1b & 5000 & 1  & $3 \times 10^{-4}$ & 2 & 12.2853 & 14.2225 \\
A1b & 5000 & 1  & $3 \times 10^{-4}$ & 8 & 12.0728 & 14.2225 \\
A1b & 5000 & 1   & $3 \times 10^{-4}$ & 16  & 11.9547  & 14.2225 \\
A1b & 5000 & 1   & $3 \times 10^{-4}$ & 32  & 12.3642  & 14.2225  \\
A1b & 5000 & 1   & $3 \times 10^{-4}$ & 128  & 12.3058  & 14.2225  \\
A1c & 5000 & 1   & $3 \times 10^{-4}$ & 8  & 12.0513  & 15.5002  \\
A1c & 5000 & 3   & $3 \times 10^{-4}$ & 8  & 12.2093  & 15.5002 \\
A1c & 5000 & 5   & $3 \times 10^{-4}$ & 8  & 11.8294  & 15.5002  \\
A1c & 5000 & 7   & $3 \times 10^{-4}$ & 8  & 12.3678  & 15.5002 \\
A1c & 5000 & 9   & $3 \times 10^{-4}$ & 8  & 12.6696  & 15.5002 \\
A1c & 5000 & 11   & $3 \times 10^{-4}$ & 8  & 14.2772  & 15.5002 \\
A1c & 5000 & 13   & $3 \times 10^{-4}$ & 8  & 12.0217  & 15.5002\\
A1d & 5000 & 1   & $1 \times 10^{-5}$ & 16  & 12.6422  & 14.4045\\
A1d & 5000 & 1   & $1 \times 10^{-4}$ & 16  & 12.3374  & 14.4045\\
A1d & 5000 & 1   & $3 \times 10^{-4}$ & 16  & 12.0456  & 14.4045\\
A1d & 5000 & 1   & $1 \times 10^{-3}$ & 16  & 12.3499  & 14.4045\\
A1d & 5000 & 1   & $3 \times 10^{-3}$ & 16  & 13.1506  & 14.4045\\
\botrule
\end{tabular*}

\footnotetext{Note: *Supplementary Figure~\ref{figS1} contains the full dataset, while Fig.~\ref{fig2} presents selected representative data due to space limitations. The information listed in this table therefore corresponds to Supplementary Figure~\ref{figS1} and includes all data shown in Fig.~\ref{fig2}.}
\end{sidewaystable}

\newpage

\begin{sidewaystable}
\caption{Experimental parameters used for the scattering correction in Fig.~\ref{fig3} and Fig.~\ref{fig4} }\label{TableS2}
\footnotesize
\setlength{\tabcolsep}{3pt}
\begin{tabular*}{\textheight}{@{\extracolsep{\fill}}cccccccccccc@{}}
\toprule
Fig.
 & \begin{tabular}[b]{@{}c@{}}Depth\\($\mu$m)\end{tabular}
 & \begin{tabular}[b]{@{}c@{}}Power\textsuperscript{1}\\(mW)\end{tabular}
 & \begin{tabular}[b]{@{}c@{}}Rate\textsuperscript{2}\\(kHz)\end{tabular}
 & $N_m$\textsuperscript{3}
 & $N_s$\textsuperscript{4}
 & \begin{tabular}[b]{@{}c@{}}Kernel\\size\end{tabular}
 & \begin{tabular}[b]{@{}c@{}}TV\\coeff.\end{tabular}
 & \begin{tabular}[b]{@{}c@{}}Feature\\dim.\end{tabular}
 & \begin{tabular}[b]{@{}c@{}}$t_{m}$/mask\\(s)\textsuperscript{6}\end{tabular}
 & \begin{tabular}[b]{@{}c@{}}$t_{c}$/mask\\(s)\textsuperscript{7}\end{tabular}
 & Sample\\
\midrule
3c & 710 & 22.74 & 0.5 & 5000 & 9 & 1 & $3 \times 10^{-4}$ & 16 & 16.92 & 6.64 & FITC C57BL/6J\\
3c & 830 & 46.05 & 0.5 & 5000 & 6 & 1 & $3 \times 10^{-4}$ & 16 & 18.05 & 6.74 & FITC C57BL/6J\\
3c & 920 & 78.20 & 0.5 & 5000 & 9 & 1 & $3 \times 10^{-4}$ & 16 & 17.08 & 6.40 & FITC C57BL/6J\\
3c & 1010 & 100 & 0.5 & 5000 & 6 & 1 & $3 \times 10^{-4}$ & 16 & 16.73 & 6.47 & FITC C57BL/6J\\
3c & 1060 & 100 & 0.5 & 5000 & 6 & 1 & $3 \times 10^{-4}$ & 16 & 16.30 & 6.74 & FITC C57BL/6J\\
4c & 730 & 24.42 & 2 & 5000 & 11 & 1 & $3 \times 10^{-4}$ & 16 & 5.02 & 5.91 & Thy1-YFP-H\\
4c & 800 & 36.87 & 2 & 5000 & 8 & 1 & $3 \times 10^{-4}$ & 16 & 5.40 & 6.19 & Thy1-YFP-H\\
4c & 870 & 55.65 & 2 & 5000 & 9 & 1 & $3 \times 10^{-4}$ & 16 & 5.13 & 6.06 & Thy1-YFP-H\\
4c & 1020 & 66.67 & 1 & 5000 & 10 & 1 & $3 \times 10^{-4}$ & 16 & 9.30 & 5.94 & Thy1-YFP-H\\
4c & 1180 & 100 & 1 & 5000 & 13 & 1 & $3 \times 10^{-4}$ & 16 & 8.91 & 5.79 & Thy1-YFP-H\\
\botrule
\end{tabular*}
\end{sidewaystable}

\end{appendices}


\end{document}